\documentclass[aps,twocolumn,pre,superscriptaddress,notitlepage]{revtex4-1}
\usepackage{epsfig}
\usepackage{natbib}
\usepackage{xspace}
\usepackage{scalerel}
\usepackage{xcolor}
\usepackage{tikz}
\usepackage{upgreek}
\usetikzlibrary{arrows.meta}
\usepackage{abraces}
\usepackage{bbold}
\usepackage{mathtools}
\usepackage{array,amsmath}
\usepackage[utf8]{inputenc}
\usepackage{amsmath}
\usepackage{bm}
\usepackage{hyperref}
\usepackage{lipsum}
\usepackage[normalem]{ulem}
 \usepackage{nicematrix}

 \newcommand{\Oin}{\Omega^{\textrm{in}}}
 \newcommand{\Oout}{\Omega^{\textrm{out}}}
 \newcommand{\Ocore}{\Omega^{c}}

\begin{document}

\title{
Decoding the interaction mediators from landscape-induced spatial patterns
}

\author{E. H.  Colombo}\email[Corresponding author: ]{e.colombo@hzdr.de}
\affiliation{Center for Advanced Systems Understanding, Untermarkt 20, 02826 G\"{o}rlitz\looseness=-1}
\affiliation{Helmholtz-Zentrum Dresden-Rossendorf, Bautzner Landstra{\ss}e 400, 01328 Dresden\looseness=-1}

\author{L. Defaveri}
\affiliation{Department of Physics, Bar Ilan University, Ramat-Gan 52900, Israel\looseness=-1}

\author{C. Anteneodo}
\affiliation{Department of Physics, PUC-Rio, Rua Marqu\^es de S\~ao Vicente  225, 22451-900 G\'avea, Rio de Janeiro, Brazil}
\affiliation{National Institute of Science and Technology for Complex Systems, 22290-180, Rio de Janeiro, Brazil\looseness=-1}

\begin{abstract}
Interactions between organisms are mediated by an intricate network of physico-chemical substances and other organisms. 
Understanding the dynamics of mediators and how they shape the population spatial distribution is key to predict ecological outcomes and how they would be transformed by changes in environmental constraints. However, due to the inherent complexity involved, this task is often unfeasible, from the empirical and theoretical perspectives.
In this paper, we make progress in addressing this central issue, creating a bridge that provides a two-way connection between the  features of the ensemble of underlying mediators and the wrinkles in the population density induced by a landscape defect (or spatial perturbation). 
The bridge is constructed by applying the Feynman-Vernon decomposition, which disentangles the influences among the focal population and the mediators in a compact way. 
This is achieved though an interaction kernel, which effectively incorporates the mediators' degrees of freedom, explaining the emergence of nonlocal influence between individuals, an ad hoc assumption in modeling population dynamics. 
 Concrete examples are worked out and reveal the complexity behind a possible top-down inference procedure.  
\end{abstract}

\maketitle

\section{Introduction}
 Organisms, at the population level, often exhibit  spatial patterns. 
These can be a spontaneous result that emerges from the interactions~\cite{camazine2020self} or forced by environmental stresses~\cite{page2003pattern,ridolfi2011noise,echeverria2023effect}, such as landscape defects that abruptly shift environmental conditions~\cite{turner2001landscape}. 
Across scales, these patterns have shown to control key ecological outcomes~\cite{rietkerk_self-organized_2004}, for example, enhancing population stability and resilience, as observed in vegetation cover~\cite{rietkerk,bonachela_termite_2015,tarnita2017} and mussel beds~\cite{koppel2005scale,koppel008,zhao_fairy_2021}.

Along with its ecological relevance, pattern complex structure can encode information about the underlying dynamics responsible for mediating the interactions between individuals (activators and inhibitors), which is typically hidden from remote sensing approaches. Decoding this complex structure could then reveal valuable information about the mechanisms behind pattern formation~\cite{zhao2020learning,ruiz2017,Edelstein,dornelas2021landscape}.
This could, at low cost,  boost the information acquired from satellite images, which is capable of generating abundant spatio-temporal data, e.g., clearly showing the distribution of vegetation~\cite{tarnita2017}, but likely miss other actors that indirectly shape plant-plant interactions, especially those below ground (e.g., water,  nitrogen, phosphorus, potassium, et al.)~\cite{MARTINEZGARCIA2023,reviewPlantComm}. A similar situation can occur at the sea, where phytoplankton is easily tracked~\cite{steele_spatial_1978}, but zooplankton, viruses and other organisms that affect phytoplankton-phytoplankton interactions remain invisible, thereby requiring expensive large-scale collaborations to gather in-situ samples~\cite{mackas1979spectral,laber2018coccolithovirus}. 

Nevertheless, a precise connection between patterns (large scales) and the underlying microscopic dynamics (small scales) is often not feasible far from equilibrium, due to the presence of strong nonlinearities~\cite{anderson1972more}.
In this work, we investigate a scenario where the effects of nonlinearities are attenuated, under a near-equilibrium scenario, opening the possibility of establishing such a connection.
Namely, we exploit the wrinkles in the population distribution induced by a landscape defect to reveal information about the set of interaction mediators from large-scale observations.
This is because, defects, by exciting a wide range of scales, put in evidence the difference in response to each scale~\cite{ridolfi2011noise,horsthemke2006noise}, which can then be associated to the dynamics of a specific mediator~\cite{page2003pattern,dornelas2021landscape}.

\begin{figure*}[t!]
    \centering
   \includegraphics[width=2\columnwidth]{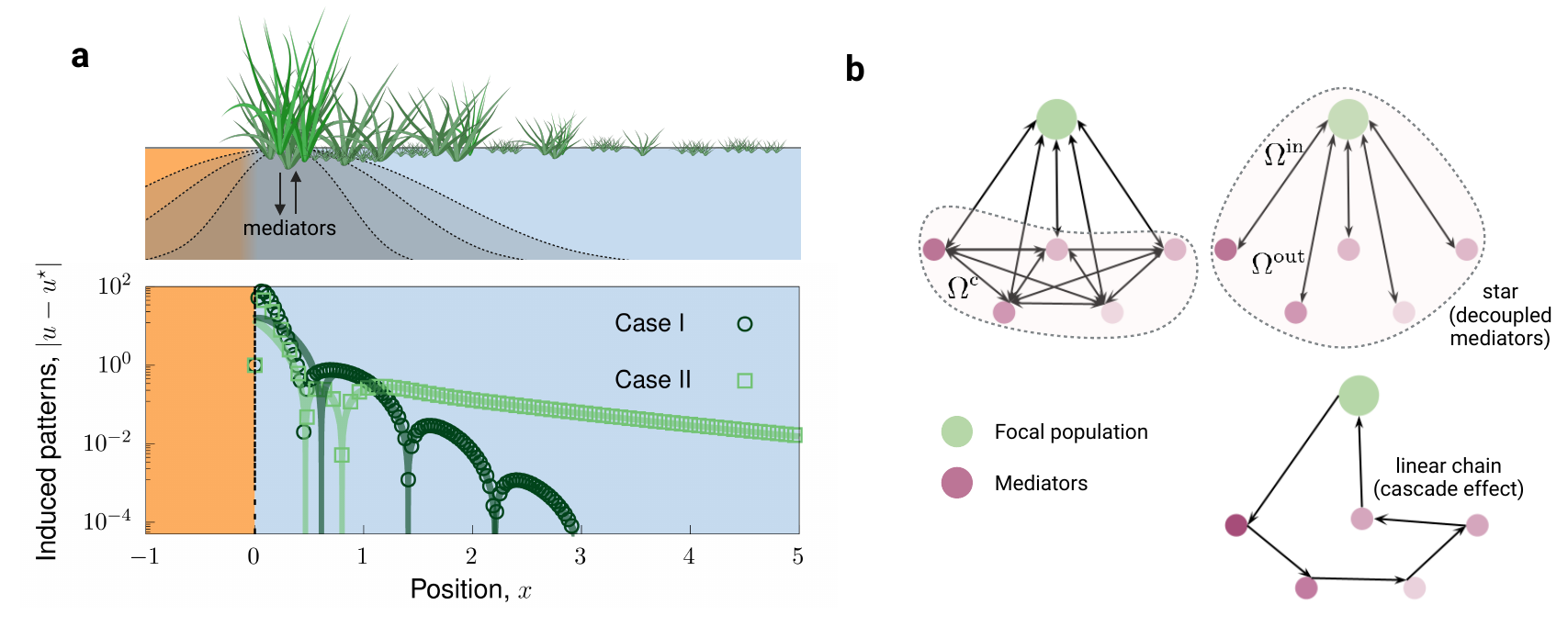} 
    \caption{\textbf{Pictorial setting, interaction network and induced patterns.} (a) Top panel shows an illustrative scheme motivated by observed vegetation and seagrass meadow patterns~\cite{juergens,ruiz2017}. It shows the focal population (vegetation), many underground mediators that extend up to different ranges, and the environmental abrupt change (from orange/adverse to blue/favorable). Bottom panel shows the results of our calculations, displaying the induced patterns close to the habitat boundary at $x=0$. Case I: one inhibiting mediator with $D_1=1$, and Case II: three inhibiting mediators with $\{D_1,D_2,D_3\}=\{0.25,1,4\}$, all having decay rate $\mu=1$ (see \ref{app:sec:examples}~for details). In both cases, we plot the long-time numerical solution  (symbols) and the theoretical prediction (solid lines), as explained in Sec.~\ref{sec:applications} (see  \textit{Inhibiting mediator} topic). 
     (b) Mediating interaction structures (all-to-all, star and linear chain), where $u$ is the focal population density and $\{v_i\}$   are the densities of the $N$ mediators which can be interconnected. The coupling scheme is given by a matrix $\Omega$ whose elements $\Omega_{ij}$ are the rates at which $i$ facilitates the production of $j$. In addition, we highlight relevant links: i) the interactions among mediators ($\Omega^c$); ii) the influence of the mediators on the population ($\Omega^{in}$) and; iii)  the influence of the population on the mediators ($\Omega^{out}$).
    }
    \label{fig:induced}
\end{figure*}

To access this information, we apply the Feynman-Vernon decomposition~\cite{FEYNMAN1963118} that disentangles the influences among the focal population and the mediators in a compact way.
The practical outcome is the construction of a two-way bridge between the emergent distance-dependent interactions and the network of mediators behind it (Sec.~\ref{sec:effective}). 
For concrete examples, in Sec.~\ref{sec:applications}, starting from an observed induced pattern~(see Fig.~\ref{fig:induced}a), we obtain the effective distance-dependent interaction kernel, and reciprocally from it we  recover information about the mediators. 
Final remarks can be found in Sec.~\ref{sec:final}.

\section{Induced states}
\label{sec:explicit}
 
 We are interested in scenarios where the population density is spatially uniform when the environment is homogeneous, i.e.,   patterns are not spontaneously formed. But, the occurrence of defects can 
 induce spatial variability of the population density, in such a way that spatial scales, which are hidden in the  absence of defects, would become revealed~\cite{horsthemke2006noise}. In Fig.~\ref{fig:induced}, we pictorially represent this scenario inspired by the vegetation spatial distribution close to a termite mount~\cite{juergens}. 

The starting point is an explicit mathematical description  of the nonlinear dynamics of 
the state vector $\psi=[u,v_1,\ldots,v_{N}]$, whose components represent 
the spatial distribution, in one dimension, of the focal population ($u(x,t)$) and of the $N$ mediators ($v_i(x,t)$)~\cite{rietkerk2008regular,cross1993pattern}. 
A general model for this dynamics is represented by  Eq.~(\ref{eq:ap_main_model}). 
See Appendix \ref{app:sec:explicit} for mathematical details. 
This description accounts for general  density-dependent rates of diffusion and growth, as well  as for the influence of an external constraint, $q(x)$, which represents the environment heterogeneity that affects the focal population, and can induce patterns, as  depicted in Fig.~\ref{fig:induced}a for the case of an abrupt change from nonviable (orange) to viable (blue) regions.

The set of differential equations that describe the dynamics is then linearized around the homogeneous steady state $\psi^\star=[u^{\star},v_1^{\star},v_2^{\star}, \ldots, v_{N}^\star]$, by considering  $\psi = \psi^\star+ \epsilon$, where   $\epsilon = [\epsilon_0,\epsilon_1,\epsilon_2,\cdots ]$ is a small deviation. Then, the linearized model, up to first order in $\epsilon$ and $q$,  is given by
\begin{eqnarray}\label{eq:eps0}  
		\partial_t \epsilon_0 &=& D_0\nabla^2 \epsilon_0  
  + \Omega_{00}\,\epsilon_0 
  +   \sum_{i=1}^{N} \Omega_{0i}\, \epsilon_i 
  + q(x)   \,,\\ \label{eq:epsj}  
\partial_t \epsilon_i &=& D_i\nabla^2 \epsilon_i  + \Omega_{ii}\,\epsilon_i +   \sum_{\substack{j=0 \\ j\neq i}}^{N} \Omega_{ij} \,\epsilon_j , \;\;\;\mbox{for $i\geq 1$}\, . 
	\end{eqnarray}
 The coefficients $\Omega_{ij}$  set  the extent to which  $j$ produces  $i$ and  results from the linearization of the reaction rates (Appendix \ref{app:sec:explicit}). Moreover, we assume $\Omega_{ii}<0$ for the stability of the uncoupled system.
 The coefficients $D_i$, with $i\ge 0$, set the diffusion rates. 
Particular forms of the interaction matrix with elements $\Omega_{ij}$ are provided in Fig.~\ref{fig:induced}b: all-to-all, star and linear chain structures. 

Considering that the system (\ref{eq:eps0}-\ref{eq:epsj}) is intrinsically stable (no spontaneous pattern formation occurs), the defect-induced wrinkles in the focal population density distribution will achieve a stationary form at long times~(see Fig.~\ref{fig:induced}a). For shallow defects, we find an expression for the wrinkles induced in the focal population density, $\epsilon_0(x)$, (see details in Appendix \ref{app:sec:explicit}). 
In Fourier space (flagged by the hat symbol), we obtain  
\begin{eqnarray}
\label{eq:induced_explicit}
    \widehat{\epsilon}_0(k) = 
 \frac{\widehat{Q}(k)}{-\det  {R}(k)}\, ,
\end{eqnarray}
where $\widehat{Q}(k)$
is the transformed external forcing,  
 and $R$ is the matrix given by the reaction and diffusion rates, with elements $R_{ij}=  
 \Omega_{ij}- \delta_{ij} D_i k^2$.

The Fourier-inversion of Eq.~(\ref{eq:induced_explicit}) provides the shape of the induced pattern in real space, namely
\begin{align}\label{eq:sol_poles}
    \epsilon_0(x) = \sum_{j=1}^{N+1} c_j e^{(i k_j   -  1/\ell_j) x},
\end{align}
where (as becomes clear from residue integration to calculate the inverse Fourier transform of Eq.~(\ref{eq:induced_explicit})), the oscillation parameters $k_j$ (wavenumber) and  $1/\ell_j$ (inverse of the decay-length) are the absolute values of the real and imaginary parts of the $j$-th zero of $\det R$~\cite{Borgogno2009,dornelas2021landscape}, and the constant coefficients $c_j$ depend on the perturbation $\widetilde{Q}$, which is assumed to be non-periodic (i.e., it does not add any characteristic mode by itself).

However, with this straightforward approach, any connection between the matrix $R$ and the features of the wrinkles (wavenumber and decay-length) is impractical: $\det {R}$ is an $N+1$-degree polynomial in $k^2$ with coefficients given by complicated combinations of diffusion and reaction rates (Appendix \ref{app:sec:explicit}). Then, the general solution (\ref{eq:sol_poles}) 
 does not help clarify how the mediators control the spatial pattern, nor does it help interpret what the pattern could tell us about the mediators.

\section{Disentangling the impact of mediators}
\label{sec:effective}

In order to  disentangle the contribution of   different sources to Eq.~(\ref{eq:induced_explicit}), we first proceed to identify in Eqs.~(\ref{eq:eps0})-(\ref{eq:epsj}),   the influence of the focal population on the ensemble of mediators by defining the vector $\Oout_i = \Omega_{i0}$
, and, reversely, the mediators' feedback on the focal population, $\Oin_i =\Omega_{0i}$. 
Moreover, we define the core, $\Ocore$ of the matrix $\Omega$, where the first row and first column have been eliminated, which sets the coupling between mediators. 
To separate the different contributions, we proceed to diagonalize, in Fourier space, the $N\times N$ core matrix associated to the mediators, following in essence, the Feynman-Vernon decomposition~\cite{FEYNMAN1963118} (for details see Appendix~\ref{app:sec:decomposition}).
The result allowed us to express the shape of the induced states of the focal population as 
\begin{eqnarray} \label{eq:epsilon0}
    \widehat\epsilon_0(k) = \frac{\widehat{q}(k)}{  D_0 k^2 - \Omega_{00} + \widehat{\mathcal{G}}(k)  }\, ,
\end{eqnarray}
with 
\begin{eqnarray}
    \widehat{\mathcal{G}}(k) \equiv -\sum_{i=1}^{N} \frac{A_i}{\lambda_i^2 + k^2} \,,
    \label{eq:CL_kernel_main}
\end{eqnarray}
where the weights $A_i$ and spatial frequencies $\lambda_i$ (for $i=1,\ldots,N$) encapsulate the information about mediators, being
\begin{equation} \label{eq:Ai}
A_i \equiv (\Oin P)_i  (P^{-1} D^{-1}\Oout)_i \,.
\end{equation}
Here $D$ is the diagonal matrix of the diffusivities of the mediators, with elements $\delta_{ij}D_i$, for $1\le i \le N$, and $P$ is the transformation matrix that diagonalizes  $D^{-1}\Ocore$, which carries the information on diffusion coefficients and mediator-mediator couplings.  
$P$ has columns given by  the eigenvectors, whose eigenvalues  are   $-\lambda_i^2$, such that $\lambda_i$ has units of inverse length. (In the current context, we are restricted to the case where the collective effect of the mediators produces real and negative eigenvalues). 
For instance, for the decoupled case, these eigenvalues are the ratios between minus the decay rates  and the respective diffusion coefficient, that is
$-\lambda_i^2=\Omega_{ii}/D_i$.

Note that  Eqs.~(\ref{eq:induced_explicit}) and~(\ref{eq:epsilon0}) 
 must coincide, but in the later  the different components that control the induced state are neatly separated, while they are intermingled in
 $\det R$ (see Appendix~\ref{app:sec:explicit} ).
  Importantly, the
decomposition allows us to obtain features of the ensemble of
mediators, re-writting Eq.-(\ref{eq:mediators_CL}) as
\begin{eqnarray} \label{eq:J}
    \widehat{\mathcal{G}}(k) = \int_{0}^{\infty} \frac{J(\lambda)}{\lambda^2 +k^2} d\lambda\,,
\end{eqnarray}
identifying,
\begin{eqnarray}
\label{eq:spectral}
    J(\lambda) = -\sum_{i=1}^N A_i \, \delta(\lambda - \lambda_i) \, ,
\end{eqnarray}
which stores the introduced spatial frequencies.
Taking the inverse Fourier transform of Eq.~(\ref{eq:J}), we obtain
\begin{eqnarray}
    \mathcal{G}(x) &=& \int_0^\infty  \frac{J(\lambda)}{2 \lambda }e^{- \lambda |x| } d\lambda\,,    
\end{eqnarray}
where the symmetry ($\pm x$) in $\mathcal{G}$ is consequence of (isotropic) diffusion of mediators.

We identify ${\cal G}(x)$ as an interaction kernel, which arises when compressing the degrees of freedom associated to the mediators. In real space, it acts through a convolution term, $\int_{-\infty}^{+\infty} \mathcal{G}(x-x') \epsilon_0(x') dx'$, which couples changes in densities at $x$ with the densities at $x'$, spatially-extending interactions (in the absence of mediators, the interaction is local). 
This result that explains the emergence of a nonlocal influence between individuals, an ingredient often proposed on and ad hoc basis to model population dynamics~\cite{murray2002}.

Finally, using the exponential form that arises from the linear propagation-decay dynamics of the mediators, we can write $G(x)$ by using the Laplace transform to help synthesize the two-way connection between the effective nonlocal interaction and the characterisctic scales of the ensemble of mediators, namely
\begin{eqnarray}
\label{eq:twoway-gj}
    \mathcal{G}(x) &=&   \mathcal{L}\{  J (\lambda)/(2\lambda) \}(|x|) \\ \notag
    \quad&&\textrm{ and }\\ \label{eq:twoway-jg}
    J(\lambda) &=&  2\lambda\mathcal{L}^{-1}\{\mathcal{G}(x) \}(\lambda)\, .
\end{eqnarray}
where the direct transformation goes from $\lambda \to |x|$ and the inverse from $|x| \to \lambda$.

\section{Examples and discussion}
\label{sec:applications}

A large class of models, after linearization, fall within the general structure of Eqs.~(\ref{eq:eps0}-\ref{eq:epsj}). 
When the spatially uniform solution is stable in an homogeneous environment, patterns can emerge, induced by the occurrence of spatial disturbances. 
In the following examples, we discuss cases related to vegetation dynamics, which are paradigmatic of pattern formation studies.
First, we will follow a direct path, deriving the interaction kernel, $\mathcal{G}$, from the mediator dynamics, i.e., the spectral density $J$, under different scenarios (Eq.~(\ref{eq:twoway-gj}). Second, attempting an inference procedure, we extract the interaction kernel from an observed pattern (using Eq.~\ref{eq:epsilon0}) and apply Eq.~(\ref{eq:twoway-jg}) to reveal the dynamics of the mediators, through $J$.

\noindent
\textit{Activating mediator---}Let us consider the specific  scenario of vegetation patterns in semi-arid regions. These systems have received continued attention due to their importance for ecosystem tolerance to dryness and sensitivity to climate change~\cite{rietkerk2008regular}.
In this context, water is the main important resource, and therefore, acts, as a main activating mediator of the vegetation cover dynamics~\cite{von2001diversity}. Diverse models have emerged in the literature  (see recent and past broad reviews~\cite{MARTINEZGARCIA2023,Borgogno2009}).

In Appendix~\ref{app:sec:activator}, we solve an extended version of the Klausmeier model~\cite{Klausmeier1999} applied to flat landscapes~\cite{MARTINEZGARCIA2023}. There, as plants consume water, they affect the water concentration not only at the consumption site but also in their vicinity due to water spatial dynamics. The resulting interaction kernel is derived and found to have the exponential form
\begin{equation}
    \mathcal{G}(x)\propto \exp(-\lambda|x|),
\end{equation}
where the inverse length $\lambda$ depends on the water parameters and precipitation rate.

\noindent
\textit{Inhibiting mediator---}Plant-plant interactions can also be mediated by substances that negatively affect plant growth. For example, seagrass meadows (which live underwater) have their interaction strongly mediated by sulfide which has been shown to shape their spatial arrangements~\cite{ruiz2023self}. 

This scenario falls within the structure depicted by Case I, in Fig.~\ref{fig:induced}a. Dots correspond to the numerical simulations of the system of partial differential equations (\ref{eq:case1}), at long times, and the solid line to the prediction provided by Eq.~(\ref{eq:epsilon0}). Since, Eq.~(\ref{eq:epsilon0}) is only valid in the limit of small perturbations, we considered the superposition of complex exponential functions but performed curve fitting for the amplitude and phase of the oscillations. This helps highlight that outside the limit of small perturbations our calculation still correctly predicts the frequency and decay of the induced spatial oscillations (see \cite{dornelas2021landscape} for detailed discussion).
Results for activating mediators will be similar, since both types of mediators lead to an induced state driven by the same mathematical structure, namely Eq.~(\ref{eq:eps0}).

\noindent
\textit{Multiple mediators (star network)---} Systems with many uncoupled mediators (star network in Fig.~\ref{fig:induced}b), for the case of inhibitors, have been analyzed in Appendix \ref{app:sec:inhibitor}.  
In particular, we worked the  numerical simulation for the case with three inhibiting mediators (Case II), presented in Fig.~\ref{fig:induced}a. As for Case I (one inhibiting mediator), dots correspond to the numerical simulation of the system (\ref{eq:case2}) and the solid line to the  prediction by Eq.~(\ref{eq:eps0}).
In both cases, for simplicity, we assumed that we know the reaction rates (all equal to one in the example), and that we do not have information about water diffusion coefficient, such that the kernel has the form,
\begin{equation}
    \mathcal{G} (x) \propto \frac{1}{N}\sum_{i=1}^{N} \exp(-|x|/D_i).
\end{equation} 
Interestingly, the superposition of exponential functions, can emulate a non-exponential profile, such as a power law~\cite{beck2003superstatistics}, providing an unique signature of the star structure.

\noindent
\textit{Multiple mediators (linear chain)---}Another scenario which is immediately tractable is  when $\Omega$ represents a cascade structure that forms a linear chain. In this case, there is a hierarchy of populations, such that every signal $i$  
has its production influenced only by the population $(i-1)$, 
hence $\Omega_{ij} \propto \delta_{j,i-1}$, and the focal population is affected by the last ($N$th) element of the cascade~(similar to the linear chain in Fig. \ref{fig:induced}b, except that it is closed in this example).  
The interaction kernel that results from our calculations (Appendix \ref{app:sec:cascade}) is given by a product of the Green functions of the mediators,
$ {\mathcal G} = -\Omega_{0N}\Pi_{j=1}^{N}[ \Omega_{j,j-1} \,G_j ]$. 
Then, 
we can use the fact that $G$ is   exponential (given by Eq.~(\ref{eq:app-green-funcion-d1})) to rewrite the interaction kernel as
\begin{equation}
    \mathcal{G}(x) \propto \exp 
    \Bigl[-\sum_{j=1}^{N}  \,\lambda_j |x|\Bigr] = \exp\left[-  N \langle \lambda \rangle |x|\right].
\end{equation}   
In this case, we can have access to the characteristic scale averaged over all mediators.

\noindent
\textit{The inference problem---}
In the previous examples, we obtained $\mathcal{G}$ knowing the mediator dynamics, $J$. In the following, we discuss how to obtain the mediator dynamics, $J$, from pattern observation. 
 We assume that, while observing an induced pattern, $\epsilon_0(x)$, we know a priori the landscape defect, $q(x)$, and the population parameters, $D_0$ and $\Omega_{00}$. 
 The dynamics of the mediators can then be partially recovered by first extracting the interaction kernel from Eq.~(\ref{eq:epsilon0}) and, then,
 applying Eq.~(\ref{eq:twoway-jg}) to access the spectral density $J$.

 The spectral density provides the characteristic spatial frequencies, $\lambda$ associated to mediator dynamics.  
 For the case of one mediator (Case I, for example, in Fig.~\ref{fig:induced}a), the spectral density, $J$, is a single delta-function, indicating the presence of only one characteristic scale. For multiple mediators (Case II, for example, in Fig.~\ref{fig:induced}a), many delta-function appear, indicating more than one characteristic scale. The cascade case has an ambiguity, because it leads, as in the case of one mediator, to a single characteristic scale (the average spatial scale across mediators). Supplementing additional information, the characteristic scale can ultimately provide the parameter values.

 Depending on the interaction network (Fig.~\ref{fig:induced}b), the spatial frequencies can provide more or less significant insights about the parameters of the mediators. 
For instance, complex vegetation models have  three equations, accounting for vegetation, roots, surface and underground water dynamics. In this case, as the dynamics of surface and  underground water  are coupled, due to infiltration, the system structure is actually a combination of the previous examples. Consequently, the extraction of a specific mediator parameter is not trivial, as the eigenvalues will be a mixture of the mediators' parameters.

Hence, although Eq.~(\ref{eq:twoway-jg}) is an advance, containing coarse-grained information, it is still limited. The complete $\Omega$ matrix can not be recovered of course, since there is a strong degeneracy as $\mathcal{G}(k)$ can be expanded in as $N$-order polynomial (Eq.~\ref{eq:determinant}), but the interaction matrix $R$ has $N^2$ elements. In these cases, more details about the system need to be known to allow the pinpointing of the values of, for example, the  groundwater diffusion coefficient, from the observation of the patterns. In any case, Sec.~\ref{sec:effective} decomposes the explicit solution of the system Eq.~(\ref{eq:induced_explicit}), focusing on coarse-grained features of the dynamics ($A_i$ and $\lambda_i$), facilitating our understanding of how the elements of the matrices $\Omega$ and $D$ shape the induced patterns. 

\section{Final remarks}
\label{sec:final}

Induced states can occur in nature close to habitat boundaries, where there is a change in environmental conditions. 
We take advantage of the fact that the forcing of the system can reveal information about the underlying dynamics that mediates the population interactions.
As a consequence, there is an opportunity for theoretical approaches to assist remote sensing initiatives, boosting the amount of information acquired from the surface (induced) patterns, e.g., captured in satellite images.

We applied our results having in mind the case of vegetation patterns where underground dynamics play a crucial role. Roots, toxic substances, water, and termites directly and indirectly mediate plant-plant interactions, but access to their dynamics is not trivial, thereby rare in previous studies~\cite{juergens,MARTINEZGARCIA2023}. Beyond the vegetation context, small-scale synthetic experiments are also interesting opportunities as designs can be tuned to target specific ``underground'' features. For instance, in Ref.~\cite{perry2005experimental}, spatial patterns were interpreted as a signature of an unknown component of interaction between bacteria and advances in synthetic experimental populations~\cite{karig2018stochastic} can create scenarios to validate whether and how much information can be recovered just from pattern observation.

In general, our results show the benefits of exploiting naturally formed disturbances or artificial ones to help us validate and improve theoretical models. Therefore, it would be interesting to look for concrete opportunities and specialize our calculations to the corresponding scenario, e.g., extending the results to two dimensions and being calibrating the parameters with realistic values.\\

\section{Acknowledgments}

We are thankful to Emilio Hernández-García and Ricardo Martínez-García for discussions and critical reading of the manuscript.

\section{Funding}
EHC acknowledges partial funding by the Center of Advanced Systems Understanding (CASUS) which is financed by Germany’s Federal Ministry of Education and Research (BMBF). 
CA acknowledge partial financial support by the Coordenação de
Aperfeiçoamento de Pessoal de Nível Superior - Brazil (CAPES) -
Finance Code 001, 
 Conselho Nacional de Desenvolvimento Científico e Tecnológico (CNPq)-Brazil (311435/2020-3) and Fundação de Amparo à Pesquisa do
Estado de Rio de Janeiro (FAPERJ)-Brazil (CNE E-26/201.109/2021).

\bibliography{induced}

\appendix

\section{Explicit approach}
\label{app:sec:explicit}

The starting (nonlinear) model for the densities of the focal population and mediators is given by
\begin{align}
\notag
\partial_t u(x,t) &= D_0(u,\{v_i\})\nabla^2 u + f_0(u,\{v_i\}) + h(x)u,\\
\partial_t v_i(x,t) &= D_i(u,\{v_i\})\nabla^2 v_i + f_i(u,\{v_i\}) ,
\label{eq:ap_main_model}
\end{align}
with $i > 0$, 
where $h(x)$ contains the heterogeneity of the environment, $D_i$ is the diffusion coefficient of mediator $i$,  and the functions $f_i$ represent the rates of the interaction (reaction) processes.  
Let us define the state vector $\psi(x,t)=[u,v_1,\ldots,v_{N}]$ and linearize the system (\ref{eq:ap_main_model})  around the homogeneous steady state $\psi^\star$, by considering  $\psi = \psi^\star+ \epsilon$, where   $\epsilon$ is a small deviation from the uniform solution. Up to first order in $\epsilon$ and $h$,  we obtain Eq.~(\ref{eq:eps0}), 
with
\begin{eqnarray}
      \Omega_{ij} &=& \partial f_i/\partial \epsilon_j|_{\psi^\star}, 
\end{eqnarray}
    where $\partial f_i/\partial \epsilon_j|_{\psi^\star}$ span the Jacobian of $f$, computed at $\psi^\star$, which is obtained from 
$f_j(u^\star,\{v_i^\star\})=0$, $\forall j$.
The Fourier transformed linearized equation reads 
\begin{align} \label{eq:epsilon}
    \partial_t \widehat{\epsilon}(k,t) =  {R}(k^2)\, \widehat{\epsilon}(k) + \widehat{q}(k)\, ,
\end{align}
where hat means Fourier transform, being $k$  the wavenumber, $R(k^2)=\Omega-Dk^2$ is a $(N+1)\times (N+1)$ matrix  where $D$ is the diagonal matrix of diffusion coefficients. The vector $q$ is an scaled version of the perturbation, $\widehat{q}(k)\equiv[\widehat{h}( k)u^\star, 0,0, \ldots,0]_{N+1}$.  
As a consequence, the stationary distribution of the population in Fourier space is given by
\begin{align}
\widehat{\epsilon}( k) &=  - {R}^{-1}   {q}  = \frac{ [\textrm{adj}( {R})\,  \widehat{q}]}{-\det  {R} }\, ,
\label{eq:fourier_ss1}
\end{align}
from which we extract the component associated to the focal species 
\begin{eqnarray}
    \widehat\epsilon_0(k) &=& \frac{\widehat{Q}(k)}{-\det  {R}}\, ,
    \label{eq:solution_explicit}
\end{eqnarray}
where we have identified the numerator with a Fourier-transformed  external forcing   $\widehat{Q}(k)=[\textrm{adj}( {R}) \widehat{q}(k)]_0$. 
On the other hand, note that 
\begin{equation} \label{eq:detM}
    \det {R} =\det [R' -k^2I]\;\prod_{i= 0}^{N} D_i,
\end{equation}
where 
$R'_{ij}=   \Omega_{ij}/D_i$.  
Hence it is clear that  $\det R$ is a $2(N+1)$-degree polynomial in $k$, which is proportional to the characteristic polynomial of the matrix $ {R}'$,  $\det[R'-\lambda I]$,  once made the identification  $\lambda =k^2$. Then, from (\ref{eq:solution_explicit}), we have 
 \begin{eqnarray}
 \det{R} &=&   r_0 + r_2k^2 +\ldots + r_{2(N+1)}k^{2(N+1)}, 
   \label{eq:determinant}
\end{eqnarray}
where the explicit form of the coefficients  $r_{2n}$ is 
given by Eq.~(\ref{eq:m}).  

To obtain these coefficients, note that  for a matrix $A$ of order $M$ (with non zero eigenvalues), using that $ \det( \exp A)=\exp( {\rm tr}A )$, we have
\begin{eqnarray} \nonumber
&&\det[A-\lambda I] = (-1)^M\lambda^{M} \exp\bigl( {\rm tr}[\log(1-A/\lambda)] \bigr) \\ \nonumber 
&=& (-1)^M\lambda^{M} \exp\bigl( -{\rm tr}\sum_{n=1}^\infty (A/\lambda)^n/n\bigr) =\sum_{j=0}^{N} a_j \lambda^j,    
\end{eqnarray}
where $a_M=(-1)^M$ and, for $0<n\le M$,
\begin{equation}
 a_{M-n}=\frac{(-1)^{M-n}}{n!} 
 \begin{vmatrix}
T_1 & n-1 & 0 & \ldots &\\
T_2 & T_1 & n-2 &  \ldots &\\
\vdots & \vdots &  &  \ddots & \vdots\\
T_{n-1} &  T_{n-2} & &  \ldots & 1\\
T_n &  T_{n-1} & &  \ldots & t_1
\end{vmatrix}, \label{eq:an}
\end{equation}
with $T_n={\rm tr}A^n=\sum_i \lambda_i^n$. 
Then,  the coefficients of the polynomial that represents $\det R$ in Eq.~(\ref{eq:detM})  
 are given by
\begin{equation} \label{eq:m}
    r_{2n}= a_n \prod_{i= 0}^{N} D_i,
\end{equation}
with the coefficients $a_n$ calculated for $A=R'$ and $M=N+1$.

\section{Disentangled representation of induced patterns}
\label{app:sec:decomposition}
 
In Eqs.~(\ref{eq:eps0})-(\ref{eq:epsj}), we first identify the influence of the focal population on the ensemble of mediators by defining the vector $\Oout_i = \Omega_{i0}$
, and, reversely, the mediators' feedback on the focal population, $\Oin_i =\Omega_{0i}$. 
Then, the core, $\Ocore$ of the matrix $\Omega$, where the first row and first column have been eliminated, sets the coupling between mediators. As a consequence,
the stationary form of Eqs.~(\ref{eq:eps0})-(\ref{eq:epsj}) can be written as
\begin{align}\label{eq:focal_CL0}
      D_0\partial_x^2 \epsilon_0(x) &+  \Omega_{00} \epsilon_0(x) + \sum_{i=1}^{N} {\Oin_i} {\epsilon}_i(x) + q(x) =0\, , \\
      D_i\partial_x^2 {\epsilon}_i(x) &+  
     %
     %
      {\Oout_i} \, \epsilon_0 (x) + \sum_{j=1}^{N}\Ocore_{ij} \epsilon_j(x)=0 \, ,     \label{eq:mediators_CL}
\end{align}
where $i=1,2,\ldots,N$.
Fourier transforming Eq.~(\ref{eq:mediators_CL}), we have
\begin{align} \label{eq:vector}
    D_ik^2\widehat{\epsilon}^{\,m}_i -\sum_{j=1}^N\Ocore_{ij}\widehat{\epsilon}^{\,m}_j = \widehat{\epsilon}_0\,\Oout_i ,
\end{align}
where we have defined the $N$-dimensional vector
$\widehat{\epsilon}^{\,m}\equiv \widehat{\epsilon}^{\,m}(k)$, whose components are the deviations of the mediators from their uniform steady states (that is, it is a reduced form of the vector $\widehat{\epsilon}$, where the first element $\widehat{\epsilon}_0$) has been removed. 
Defining the diagonal matrix $D$ of elements $\delta_{ij}D_i$, with $1\le i \le N$, we rewrite Eq.~(\ref{eq:vector}) in matrix form as
\begin{align} \label{eq:matrix}
   ( k^2 I_N  -D^{-1}\Ocore )\widehat{\epsilon}^{\,m}  = \widehat{\epsilon}_0 D^{-1}\Oout ,
\end{align}
where $I_N$ is the $N\times N$ identity matrix, and
recalling that $\Ocore$ is a matrix, $\Oout$ and $\Oin$ are vectors, while $\widehat{\epsilon}_0$ is a scalar. 
Then, from Eq.~(\ref{eq:focal_CL0}), we 
obtain the solution  

\begin{align}
    \widehat{\epsilon}^{\, m}(k) 
    &= \widehat{\epsilon}_0(k)\,[k^2 I_N   - D^{-1}\Ocore]^{-1} 
    D^{-1}\Oout \\
    &= \widehat{\epsilon}_0(k)\,[ P  \,\Lambda(k) \, P^{-1}] \,  
    D^{-1}\Oout ,
    \label{eq:CL_step1}
\end{align}
where the matrix $P$ has columns given by  the eigenvectors of $D^{-1}\Ocore$   
and $\Lambda(k)$ is a diagonal matrix with elements 
$\Lambda_{ii}(k)=[k^2+\lambda_i^2]^{-1}$, where $-\lambda_i^2$ are the eigenvalues of $D^{-1}\Ocore$ (that must be negative for the stability of the system), which combine the impact of diffusion and mediator-mediator coupling.

Let us  develop the sum of the contributions of the mediators appearing in Eq.~(\ref{eq:focal_CL0}), using Eq.~(\ref{eq:CL_step1}). That is
\begin{eqnarray}
&&\sum_{i=1}^{N} {\Oin_i} {\epsilon}_i(x) = 
\Oin \cdot \widehat{\epsilon}^{\,m}(k)\\ 
&=& \widehat{\epsilon}_0(k)[ \Oin P] \,\Lambda(k) \, [P^{-1} D^{-1}\Oout]  \\
&=& \widehat{\epsilon}_0(k)\sum_{i=1}^{N} \frac{A_i}{\lambda_i^2+k^2}\,,
\label{eq:CL_eigensum}
\end{eqnarray}
which is a sum of Lorentzian functions, where $A_i \equiv (\Oin P)_i  (P^{-1} D^{-1}\Oout)_i$ weights the impact of each eigenvalue. 
 The weights depend on the eigenvectors of the core matrix (contained in $P$) and hence on how they reflect the feedback loop between the focal population and the mediators (i.e., $\Oin$ and $\Oout$), moderated by the diffusivities (contained in $D$).

Finally, into  the Fourier transformed Eq.~(\ref{eq:focal_CL0}), 
\begin{equation}
  [D_0 k^2 - \Omega_{00}]\widehat{\epsilon}_0(k) 
   + \Oin  \cdot \widehat{\epsilon}^{\,m}(k) = \widehat{q}(k)\,,   
\end{equation}
 we plug Eq.~(\ref{eq:CL_eigensum}), extracting the effective representation of the induced states of the focal population,
\begin{eqnarray} \label{eq:app-effective}
    [ D_0 k^2 - \Omega_{00} + \widehat{\mathcal{G}}(k) ]\,\widehat\epsilon_0(k) = \widehat{q}(k)\, ,
\end{eqnarray}
where we identify
\begin{eqnarray}
    \widehat{\mathcal{G}}(k) \equiv \sum_{i=1}^{N} \frac{-A_i}{\lambda_i^2 + k^2} 
    \label{eq:CL_kernel}
\end{eqnarray}
 as the interaction kernel that arises when compressing the degrees of freedom from the mediators. 
 This is because, in real space, Eq.~(\ref{eq:CL_kernel})
 becomes
\begin{eqnarray} \label{eq:app-effective-real}
    [ D_0\partial_x^2 + \Omega_{00} -  \mathcal{G} \ast ]\, \epsilon_0(x) = -{q}(x)\, ,
\end{eqnarray}
 and the kernel 
 generates a convolution term, $\mathcal{G}\ast\epsilon_0= \int_{-\infty}^{+\infty} \mathcal{G}(x-x') \epsilon_0(x') dx'$, which couples changes in densities at $x$ with the densities at $x'$, spatially-extending interactions.

\section{  Examples }
\label{app:sec:examples}

\subsection{Activator networks}
\label{app:sec:activator}

 We consider a structure that is common in models that describe vegetation in semi-arid regions, where water-type mediators act  as activators~\cite{MARTINEZGARCIA2023}. 
 As example, for the simplest case of one mediator, a typical structure is

 \begin{eqnarray}\notag
    \partial_t u(x,t) &=& D_0\partial_{xx} u  + u^2v - u  + h(x)\, ,
    \\  
     \partial_t v(x,t) &=&   \partial_{xx} v   + R -v u^2 - v,
     \label{eq:water}
\end{eqnarray}
where $R$ is the precipitation rate.
For this system, a vegetated steady state ($u^\star=R/2+\sqrt{R^2/4-1}$), and $v^\star=1/u^\star$) is stable for any $R>2$. The interaction matrix is
\begin{equation}
 \Omega=
 \begin{pNiceArray}{c|c} 
\Omega_{00}  & \Oin \\
\hline
\Oout & \Ocore 
\end{pNiceArray}
=
 \begin{pNiceArray}{c|c} 
-1  &   {u^\star}^2 \\
\hline
-2 &   -1-{u^\star}^2 \\
\end{pNiceArray},
\end{equation} 
then $A_1=\Oout \Oin =-{u^\star}^2$, the single eigenvalue is
$-\lambda^2=1+{u^\star}^2$, 
and Eq.~(\ref{eq:CL_kernel_main}) becomes
\begin{eqnarray}
    \widehat{\mathcal{G}}(k) = \frac{2{u^\star}^2}{1+{u^\star}^2+k^2}  \,,\label{eq:kernel_activators}
\end{eqnarray}
implying exponential kernel $\mathcal{G}(x)\sim \exp(-\sqrt{1+{u^\star}^2}|x|)$.

\subsection{Inhibitory networks (case studies)}
\label{app:sec:inhibitor}

 We analyze a type of inhibitory network for which numerical examples are  depicted in Fig.~\ref{fig:induced}a, where  agreement between the explicit and effective descriptions is observed.
In all cases we consider an heterogeneous scenario
with a semi-infinite habitat condition, 
where,  for $x<0$, $h\to-\infty$, mimicking harmful conditions, and for $0<x<L$, $h=0$. 
The system size, $L$, is taken to be as large as possible to account for the spatial relaxation of the induced states.

The evolution equations for $N$ inhibitors are 
of the form
\begin{eqnarray}\notag
    \partial_t u(x,t) &=& D_0\partial_{xx} u  + u - u(v_1 +\cdots + v_N) + h(x)\, ,
    \\  
     \partial_t v_i(x,t) &=&  D_i\partial_{xx} v_i  -v_i + u \;\;\;\;  \mbox{for $1\le i \le N$},
     \label{eq:caseN}
\end{eqnarray}

The homogeneous steady state in the viable habitat is given by $u^\star=v_i^\star=1/N$ for any $i$. 
  Moreover, the (linearized) network is of star type (without mediator-mediator interactions), given  by
 
 \begin{equation}
 \Omega=
 \begin{pNiceArray}{c|c} 
\Omega_{00} & \Oin \\
\hline
\Oout & \Ocore 
\end{pNiceArray}
=
 \begin{pNiceArray}{c|rrrr} 
0 & -\frac{1}{N} & -\frac{1}{N} & \cdots   & -\frac{1}{N} \\
\hline
1 & -1  &   0  & \cdots &  0 \\
1 &   0  &-1  & \cdots &  0 \\
\vdots & \vdots     & 0  & \ddots &  0 \\
1 &  0  & 0  & \cdots &  -1 \\
\end{pNiceArray}.
\end{equation}

For all $i\ge 1$,  $\Oin_i=\Omega_{0i}=-1/N$, $\Oout_i=\Omega_{i0}=1$,  $\Omega_{00}=0$,  
and the core $\Ocore$ is minus the unitary matrix (since $\Omega_{ii}=-1$), so  that $D^{-1}\Ocore$ is a diagonal $N\times N$ matrix with elements $-1/D_i$ (which are the eigenvalues $-\lambda_i^2$). 

Moreover, in this case $A_i=-1/(N D_i)$, 
hence, the interaction kernel in Eq.~(\ref{eq:CL_kernel_main}) becomes
\begin{eqnarray}
    \widehat{\mathcal{G}}(k) = \frac{1}{N}\sum_{i=1}^{N} \frac{D_i}{1+ D_i^2k^2} \,,\label{eq:kernel_inibitors}
\end{eqnarray}
and

\begin{eqnarray} \label{eq:J-inhibit}
  J(\lambda) =  \sum_{i=1}^{N} \frac{1}{ND_i}\delta(\lambda-\lambda_i)\, .
\end{eqnarray}

Then, the spectral density $J$ recovers the distribution of values of the diffusivities in the original systems.

 The case studies depicted in Fig~\ref{fig:induced} correspond to particular values of $N$ , where  
we set  $D_0=10^{-3}$.

\begin{itemize}
    \item 
 Case I: A single inhibitor ($N=1$). 
 The dynamics of the focal population and the inhibitor are explicitly described by
\begin{eqnarray}\notag
    \partial_t u(x,t) &=& D_0\partial_{xx} u  + u - vu +h(x)u,\\
     \partial_t v(x,t) &=& \partial_{xx} v  -v + u.
      \label{eq:case1}
\end{eqnarray}
\item Case II: three inhibitors ($N=3$) 
The evolution equations explicitly are 
\begin{eqnarray}\notag
    \partial_t u(x,t) &=& D_0\partial_{xx} u  + u - (v_1 + v_2 + v_3)u + h(x)\, ,\\ \notag
     \partial_t v_1(x,t) &=&  \partial_{xx} v_1  -v_1 + u \, ,\\ 
     \partial_t v_2(x,t) &=& 4\, \partial_{xx} v_2  -v_2 + u\, \\\notag
     \partial_t v_3(x,t) &=& \frac{1}{4}\partial_{xx} v_3   -v_3 + u\,  .
     \label{eq:case2}
\end{eqnarray}
 
\end{itemize}

It is worth noting that we can reach the same result for the kernel from a Green-function approach. In fact note that each term  of  Eq.~(\ref{eq:kernel_inibitors}) corresponds to the Fourier transform of  the Green function for the operator $(D_i \nabla^2+ \Omega_{ii})$, which in one dimension  is
  \begin{eqnarray}\label{eq:app-green-funcion-d1}
        G_i(x) = \frac{1}{2\sqrt{|\Omega_{ii}|D_i}}   \exp\left(-|x| \sqrt{|\Omega_{ii}|/D_i}\right).
    \end{eqnarray}
Then, the solution of 
$  ( D_i \nabla^2+ \Omega_{ii})\epsilon_i  = -\Omega_{i0} \epsilon_0$  can be written as
	\begin{eqnarray}  
		\epsilon_i  (x) &=&  \Omega_{i0} \int_{-\infty}^{\infty} G_i(x-x')  \epsilon_0(x') dx'    
 \equiv  \Omega_{i0}\,G_i\ast\epsilon_0,\;\;\;\;\;\;
  \label{eq:zero-order}
\end{eqnarray}
    where $\ast$ represents a convolution. 
    Plugging this solution into the linearized stationary equation for $\epsilon_0$,  Eq.~(\ref{eq:eps0}),   we arrive to a single closed equation for $\epsilon_0$, namely
\begin{eqnarray} \label{eq:eps0ss}
	  (D_0\nabla^2 
  +  \Omega_{00}  -\mathcal{G}\ast) \epsilon_0
  + h(x)u^\star = 0\,, 
\end{eqnarray}
with
\begin{eqnarray}     \label{eq:kernel2} 
     \mathcal{G}(x)&=& -\sum_{i=1}^{N} \Omega_{0i} \Omega_{i0}    G_i(x).
\end{eqnarray}
   In the present example, 
   for which 
   $\Omega_{i0}=1$, $\Omega_{0i}=-1/N$, and $\Omega_{ii}=-1$ for all $i>1$,  we have
 $\mathcal{G}(x)  =  \sum_{i=1}^{N}G_i(x)/N$,  
whose Fourier transform recovers Eq.~(\ref{eq:kernel_inibitors}).

\subsection{Cascade of mediators}
\label{app:sec:cascade}

Considering the linearized system
\begin{eqnarray}
\label{eq:psij_cascade1}
		\partial_t \epsilon_0 &=& D_0\nabla^2 \epsilon_0  + \Omega_{00}\epsilon_0 +  \Omega_{0N} \, \epsilon_N + q(x)  \, ,\\ \notag
		\partial_t \epsilon_i &=& D_i \nabla^2 \epsilon_i + \Omega_{ii} \epsilon_i + \Omega_{i\;i-1} \, \epsilon_{i-1},\quad \textrm{for }i\geq 1\, , 
	\end{eqnarray}
therefore, the interaction network is \begin{equation}
 \Omega=
 \begin{pNiceArray}{c|c} 
0 & \Oin \\
\hline
\Oout & \Ocore 
\end{pNiceArray}
=
 \begin{pNiceArray}{c|cccc} 
\Omega_{00} & 0 &0 & \cdots &   \Omega_{0N} \\
\hline
\Omega_{10} & \Omega_{11} & 0  & \cdots &  0 \\
0 & \Omega_{21}  &\Omega_{22}  & \cdots &  0 \\
\vdots      &   0  &0  & \ddots &  0 \\
0 &   0  & 0  & \Omega_{N\,N-1}  &  \Omega_{NN} \\
\end{pNiceArray}.
\end{equation}

Note that the core $\Ocore$ is a lower triangular matrix (bi-diagonal), then  $D^{-1}\Ocore$ is also triangular, hence the eigenvalues coincide with the diagonal elements $D_i^{-1}\Omega_{ii}$, which are real and negative.

For a clear example, let us consider $N=3$, which is enough to visualize the recurrent structure of $P$, the matrix whose columns are the eigenvectors, namely
\begin{align}
    P = \begin{pmatrix}
\frac{ (\Omega'_{11}  -  \Omega'_{22})( \Omega'_{11}  -  \Omega'_{33} )}{ \Omega'_{32} \Omega'_{21}}  & 0 & 0 \\[2mm]
\frac{( \Omega'_{11}  -  \Omega'_{33} )}{ \Omega'_{32} } & 
\frac{( \Omega'_{22}  -  \Omega'_{33} )}{ \Omega'_{32} } & 0 \\[2mm]
1 & 1 & 1
\end{pmatrix} \, ,
\end{align}

where we have defined $\Omega'_{ij}=\Omega_{ij}/D_i$, and its inverse is

\begin{align}
    P^{-1} = \begin{pmatrix}
\frac{  \Omega'_{32} \Omega'_{21}   }{\left( \Omega'_{11}  -  \Omega'_{22}  \right) \left( \Omega'_{11}  -  \Omega'_{33} \right)} & 0 & 0 \\[2mm]
\frac{ \Omega'_{32} \Omega'_{21} }{\left( \Omega'_{22}  -  \Omega'_{11} \right) \left( \Omega'_{22}  -  \Omega'_{33} \right)} & 
\frac{ \Omega'_{32} }{ \Omega'_{22}  -  \Omega'_{33} } & 0 \\[2mm]
\frac{ \Omega'_{32} \Omega'_{21} }{\left(\Omega'_{33}  -  \Omega'_{11}  \right) \left( \Omega'_{33}  -  \Omega'_{22} \right)} &   
\frac{\Omega'_{32} }{ \Omega'_{33}  -  \Omega'_{22} } & 1 
\end{pmatrix}
 \, ,
\end{align} \\[3mm] \noindent
then, we have $(\Omega^{\mathrm{in}} P )_i =  \Omega_{0N}$ for all $i$, and
\begin{align}
    (P^{-1} D^{-1} \Omega^\mathrm{out})_i = 
    \frac{\Omega_{10}}{D_1}(P^{-1})_{i1}= 
    \frac{\Omega_{10}}{D_1} \frac{\prod_{j=1}^{N-1} \frac{\Omega_{j+1\,j}}{D_{j+1}}}{\prod_{j \neq i}^N \left( \frac{\Omega_{ii}}{D_i} - \frac{\Omega_{jj}}{D_j} \right)}\,,
\end{align}
hence, from Eq.~(\ref{eq:Ai}),
\begin{align}
    A_i =    \frac{\Omega_{10}\Omega_{0N}}{D_1} \frac{\prod_{j=1}^{N-1} \frac{\Omega_{j+1\,j}}{D_{j+1}}}{\prod_{j \neq i}^N \left( \frac{\Omega_{ii}}{D_i} - \frac{\Omega_{jj}}{D_j} \right)}\,.
\end{align}

Finally, Eq.~(\ref{eq:CL_kernel_main}) becomes
\begin{align}
    \widehat{\mathcal{G}}(k) =  \sum_{i=1}^{N} \frac{-A_i}{  k^2+\frac{-\Omega_{ii}}{D_i}   } =  
    -\frac{\Omega_{0N} \prod_{j=1}^{N} \Omega_{j\,j-1}}{\prod_{j=1}^{N} 
    \left[  D_i (k^2 + \frac{-\Omega_{ii}}{D_i})\right]} \,,
\label{eq:kernel_cascade}
\end{align}
where we used Lagrange interpolation decomposition  $\frac{R(x)}{Q(x)}=\sum_{i=1}^N \frac{R(x_i)}{(x-x_i)Q'(x_i)}$,  setting $R(x)=1$, $Q(x)=\Pi_{i=1}^N(x-\Omega'_{ii})$ and $x=k^2$. 

From the Green-function perspective, Eq.~(\ref{eq:kernel_cascade}) can be written as 
\begin{eqnarray}
    \widehat{\mathcal{G}}(k) =  -\Omega_{0N}\Pi_{j=1}^{N}   [ \Omega_{j,j-1} \,G_j(k) ]\,,
    \label{eq:kernel_cascade2}
\end{eqnarray}
 where the Green function $G_j$ is the same defined in Eq.~(\ref{eq:app-green-funcion-d1}). 
In fact, in the present problem,     the stationary form of Eq.~(\ref{eq:psij_cascade1}) for 
$\epsilon_i$ is
\begin{eqnarray}
     (\Omega_{ii} + D_i \nabla^2) \epsilon_i = -\Omega_{i\;i-1} \epsilon_{i-1} \, .
\end{eqnarray}
Using the Green function $G_i$ of the operator on the left-hand side, given by Eq.~(\ref{eq:app-green-funcion-d1}), we can write $\epsilon_i$ explicitly as
\begin{eqnarray}
    \epsilon_i = \Omega_{i,i-1} G_i \ast \epsilon_{i-1} \, .
\end{eqnarray}
The first two solutions are $\epsilon_1 = \Omega_{1,0} G_1\ast \epsilon_0$, and $\epsilon_2 = \Omega_{2,1} G_2 \ast \epsilon_1 =  \Omega_{2,1}  \Omega_{1,0} G_2 \ast (G_1 \ast \epsilon_0)$, and so on recursively, such that the $i$-th term is
\begin{eqnarray} 
\label{eq:epsilon-i}
\epsilon_i ({ x}) &=&  \Pi_{j=1}^{i} 
 [\ \Omega_{j,j-1}  \,G_j \ast] \epsilon_0 (x) \,.
\end{eqnarray}

Substituting the solution for $\epsilon_N$ into Eq.~(\ref{eq:psij_cascade1}) for $\epsilon_0$, we obtain the closed equation for $\epsilon_0$
\begin{eqnarray}
\label{eq:psij_cascade2}
		( D_0\nabla^2    + \Omega_{00}\  - \;\mathcal{G}\ast )\epsilon_0  + q(x) &=& 0 \, ,
	\end{eqnarray}
where, from Eq.~(\ref{eq:epsilon-i}), the interaction kernel $\mathcal{G}$ for the cascade is 
\begin{eqnarray}
   \mathcal{G}(x)  =  -\Omega_{0N}\Pi_{j=1}^{N}   \Omega_{j,j-1} 
\;[G_1\ast G_2\ast \cdots \ast G_N](x)  \,,\;\;\;\;
\end{eqnarray}
whose  Fourier-transform recovers Eq.~(\ref{eq:kernel_cascade2}).

 \end{document}